\documentclass[proof]{WileyASNA-v1}

\articletype{Article Type}%

\received{26 April 2016}
\revised{6 June 2016}
\accepted{6 June 2016}

\begin{document}

\title{On the possibilities of classical nova identifications among historical Far Eastern guest star observations}

\author[1]{Nikolaus Vogt}

\author[2]{Susanne M Hoffmann}

\author[1]{Claus Tappert}

\authormark{Nikolaus Vogt \textsc{et al}}

\address[1]{\orgdiv{Instituto de Física y Astronomía}, \orgname{Universidad de Valparaíso}, \orgaddress{\country{Chile}}}

\address[2]{\orgdiv{Physikalisch-Astronomische Fakultät}, \orgname{Friedrich-Schiller-Universität Jena}, \orgaddress{\country{Germany}}}

\corres{Nikolaus Vogt \email{nikolaus.vogt@uv.cl}}

\presentaddress{This is sample for present Address text this is sample for present Address text}

\abstract{More than 100 guest star observations have been obtained by Chinese, Korean, Japanese and Vietnamese astronomers between $\sim 600$ BCE and $\sim 1690$ CE. Comparing the coordinates from the information given in old texts for eight supernova recoveries with modern supernova remnant positions, we estimate a typical positional accuracy of the order of 0.3$^\circ$ to $7^\circ$ for these supernovae. These values turn out to be also a start for the expected deviation angle between a classical nova observed as a guest star and its modern counterpart among known cataclysmic variables (CVs). However, there are considerable disagreements among modern authors in the interpretation of ancient Far Eastern texts, emphasizing the need to consult again the original historic sources, in order to improve the positioning reliability. We also discuss the typical amplitudes of well observed classical novae and find that modern counterparts of nova guest stars should be $V = 18$~mag and thus easily observable. In this context we also consider the ``hibernation scenario'' and conclude that it is impossible to decide from currently-available observations whether hibernation is common. In addition to the limiting magnitude around 2~mag for ancient guest star detections mentioned in the literature, we consider the possibility that also fainter guest stars (4--5~mag) could have been detected by ancient observers and give arguments in favor of this possibility. For these limits we compare the expected nova detection rate of ancient naked-eye observers with that during modern times, and conclude that they coincide in order of magnitude, which implies that, indeed, a considerable number of classical nova remnants should be hidden among the Far Eastern guest star reports. Finally, we present a statistical analysis of the probability of casual misidentifications based on frequency and galactic distribution of CVs in the AAVSO-VSX catalogue.
}

\keywords{cataclysmic binaries, historical guest stars}

\jnlcitation{\cname{%
\author{Vogt, N.}, 
\author{S. M. Hoffmann}, and 
\author{C. Tappert}} (\cyear{2019}), 
\ctitle{On the possibilities of classical nova identifications among historical Far Eastern guest star observations}, \cjournal{A.N.}, \cvol{2019; : }.}


\maketitle


\section{Introduction}\label{sec1}
The total time available for any systematic study of stellar long-term variability is rather limited: It was only in the late 19th century that the use of photographic plates permitted storing information on stellar variability in a permanent and systematic way. Therefore, our sky patrol archives cover up to 140 years of the visible sky history. Yet, in most cases the time intervals covered are much shorter for a given sky position or magnitude limit. On the other hand, there are interesting astrophysical questions requiring longer time spans: for instance the unsolved question whether or not all cataclysmic variables (CVs) regularly experience nova eruptions. 

Such an eruption in a CV system is an explosive runaway fusion of hydrogen to helium on the surface of the white dwarf primary component of a CV, after accumulation of hydrogen rich matter due to mass transfer from the secondary star. This process does not destroy the binary system, and mass transfer from the secondary star to the white dwarf recommences within a couple of years \citep{retter1998}. According to theoretical models, it repeats many times during the life of a given classical nova system at typical time intervals between $10^4$ and $10^5$ years \citep{yaron2005}. The question arises: Does the nova eruption determine the behavior of the later post-nova? For example, does it induce the transition of a low mass-transfer dwarf nova to a high mass-transfer nova-like CV (or vice versa)? These sub-types of CVs have similar binary configurations, orbital period values and many other parameters. Why does their outburst behavior differ so much? \cite{vogt1982} was the first suggesting that different sub-types of CVs could be transition stages in a large cycle between two subsequent nova eruptions. \cite{shara1986} modified this picture, introducing a ``hibernation'' stage in which, as a consequence of the nova eruption, the Roche-lobe of the secondary CV component is disconnected for some time from its surface, implying a break in the mass transfer. This question still is under debate. A possible approach to solve this puzzle is to identify CVs whose nova eruption was registered during previous millennia, i.\,e. in pre-telescopic epochs, covered by Far Eastern observations of the night sky. This would increase the total time interval of post-nova observations from currently about 500 years by a factor of $\approx 5$, thus reaching a significant fraction of the total time which passes between two subsequent nova eruptions. The aim of this paper is to analyze the reliability of previous attempts of supernova and classical nova identification (Section \ref{sec2}), to estimate the expected number of possible identifications of Far Eastern records with classical novae (Sections \ref{sec3} and \ref{sec4}) and to determine the probability of chance coincidences with CVs and other possible variable stars (Section \ref{sec5}).

\section{How reliable are ancient Far Eastern guest star observations?}\label{sec2}
In order to obtain reliable estimates on the accuracy expected for future identification efforts we give here the angular separation $\Delta\alpha$ between Far Eastern guest star positions of supernovae and their modern counterparts (pulsars and/or the central coordinates of supernova remnant nebulae). The comparison was made in galactic coordinates. For the modern counterparts of old supernovae, their coordinates given in the SIMBAD data base were used. The corresponding guest star positions are those listed by \cite{stephenson1976} and by \cite{nickiforov2010}. Table \ref{tab1} contains the results. According to this, \cite{stephenson1976} gives reasonable results for supernovae, with position accuracies of the order of angular separations $0.3^\circ \leq \Delta\alpha \leq 6.8^\circ$. On the other hand, a comparison with Nickiforov's (2010) data reveals very scattered angular separations with an extreme value of more than $26^\circ$. We consider this study as not very reliable, and do not use it anymore throughout this paper. 

 \begin{center}
\begin{table}[t]%
\centering
\caption{Angular separation values of identified supernovae.\label{tab1}}%
\tabcolsep=0pt%
\begin{tabular*}{20pc}{@{\extracolsep\fill}rcc@{\extracolsep\fill}}
\toprule
\textbf{ } & \multicolumn{2}{c}{Angular separation $\Delta\alpha$}    \\
\textbf{Supernova} & \textbf{Stephenson, 1976}  & \textbf{Nickiforov, 2010}    \\
\textbf{year CE} & \textbf{degrees}  & \textbf{degrees}    \\
\midrule
 185 & 3.22  & 1.55      \\
 386 & 6.81  & 10.46  \\
 393 & 0.30  & 4.57     \\
1006 & 2.90  & 26.69     \\
1054 & 2.81  & 1.46     \\
1181 & 3.42  & 5.55     \\
1572 & 1.17  & 5.19     \\
1604 & 1.56  & 18.73     \\
\bottomrule
\end{tabular*}
\end{table}
\end{center}

More important than the ranges of $\Delta\alpha$ are the maximum values which define the size of the search radius to be applied when identifying modern counterparts for ancient sightings. Table~\ref{tab2} lists the hitherto published attempts to identify ancient sightings with classical nova eruptions, proposing counterparts among known cataclysmic variables and/or planetary nebulae (those that were classified as such erroneously, turning out to be ancient nova shells). For the event 77 BCE there are two different identifications in the literature, one of them \citep{johansson2007} is the dwarf nova Z Cam with a remarkable shell detected by \cite{shara2007}. However, in this case we have the largest deviation $\Delta\alpha \approx 11$ degrees while the identification with the dwarf nova DO Dra (\citealt{hsi1957}; \citealt{hertzog1986}) yields a separation of only $3.7^\circ$. The identification of the event 101 CE with BK Lyn, based on Stephenson's (1976) coordinates, has also a rather large deviation; however, alternative coordinates of this event are given by \cite{hsi1957} and \cite{hertzog1986} diminishing $\Delta\alpha$ to $\approx 0.5^\circ$, a very good coincidence, confirmed by recent results of one of us (SH). This star was already recognized as varying its spectrum at short time scales by \citet{dobrzycka1992} and by \citet{szkody+howell1992}; more recently, it awoke especial interest because BK Lyn was classified as a relatively quiet nova-like star for a long time, until in 2011 it was found by \cite{patterson2013} to suddenly have switched to ER UMa-type dwarf nova outburst activity (i.\,e. SU UMa type behavior with super-outbursts and short outbursts at very short time scales). This implies a change in the mass-transfer rate, which is one of the potential long-term consequences of a nova eruption as required by the hibernation scenario \citep{shara1986}. It should be noted, however, that the records of the American Association of Variable Star Observers (AAVSO, \citet{watson2006}) show that BK Lyn since 2014 has returned to its previous quieter state.

If we consider all values in Tab.~\ref{tab2}, we get $0.5^\circ  \leq \Delta\alpha \leq 11.1^\circ$ for classical nova identifications; disregarding the large value for Z Cam, the range diminishes to $0.5^\circ \leq \Delta\alpha \leq 5.2^\circ$ which is very similar to the range found for supernovae. It seems that \cite{stephenson1976} gives also reasonable results for classical novae in most cases; his position accuracy is of the same order as that of supernovae. However, it is necessary to re-check Stephenson's positions in doubtful cases as shown by the example of BK Lyn. 

\begin{center}
\begin{table*}[t]%
\caption{Published suggestions for classical nova identifications. The angular separation refers to the difference between the guest star position and the modern counterpart, both according to the references given in the last two lines.\label{tab2}}
\centering
\begin{tabular*}{500pt}{@{\extracolsep\fill}p{18ex} p{11ex}p{11ex}p{11ex}p{11ex}p{11ex}p{11ex} @{\extracolsep\fill}}
\toprule
 &  \multicolumn{6}{c}{ \textbf{guest star year} } \\
 & \textbf{77 BCE}  & \textbf{77 BCE}  & \textbf{101 CE}   & \textbf{483 CE}  & \textbf{1437 CE}  & \textbf{1645 CE} \\ 
\midrule
Suggested modern counterpart & YY Dra (=DO Dra) & Z Cam &  BK Lyn &  PN Te11  & GDS J1701281-430612  & AT Cnc   \\\hline
angular separation $\Delta\alpha/^\circ$ & 3.7 & 11.1 & 0.5 & 4.6 & 3.3 & 5.2 \\ \hline
$P_{orb}$ (d) & 0.1674 & 0.2898 & 0.0750  & 0.12 & 0.5340 & 0.2016 \\\hline
V magnitude max. &  10 --  & 10 --   & 14.3 -- & 14.6 -- & 12.1 -- & 12.5 -- \\ 
range to min. & 17 & 14.5 & 16.5  & 17.3  &18.1  & 15.8 \\\hline
remarks on the modern counterpart &dwarf nova, intermediate polar & dwarf nova, subtype Z Cam, extended shell detected & nova-like star with dwarf nova activity & eclipsing dwarf nova, former planetary nebula & eclipsing dwarf nova, intermediate polar, former planetary nebula & dwarf nova, subtype Z Cam, extended shell detected \\\hline
reference (guest star position) & \tnote{1},\tnote{2} & \tnote{1},\tnote{2} &  \tnote{1}  &  \tnote{2}  &\tnote{2} &  \tnote{8}   \\\hline
reference (modern identification) & \tnote{3} & \tnote{4} (in reaction to \tnote{5}) & \tnote{3} & \tnote{6} &\tnote{7} &\tnote{8} \\\hline
\bottomrule
\end{tabular*}
\begin{tablenotes}
\item References in last two lines:
\item[1] \cite{hsi1957}
\item[2] \cite{stephenson1976}
\item[3] \cite{hertzog1986}
\item[4] \cite{johansson2007}
\item[5] \cite{shara2007}
\item[6] \cite{miszalski2016}
\item[7] \cite{shara2017a}
\item[8] \cite{shara2017b}
\end{tablenotes}
\end{table*}
\end{center}

\section{Are the modern counterparts of ancient classical novae too faint to be detected?}\label{sec3}
\begin{figure}[t]
	\centerline{\includegraphics[width=.95\columnwidth]{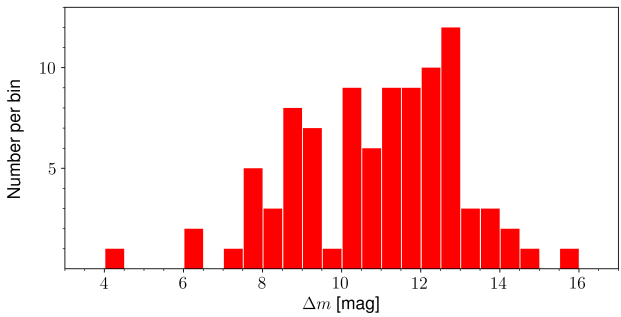}}
	\caption{Distribution of the eruption amplitudes of all modern classical novae; eruption date observations before 1986 and since 1783 (V magnitude). 
	\label{fig1}}
\end{figure}

It is relatively easy to determine the amplitude of classical novae, defined as the difference between their maximum brightness and that of the post-nova (decades after eruption). The corresponding histogram is shown in Fig.~\ref{fig1}.  We find that the vast majority has eruption amplitudes $<13.0$~mag. Naked eye observations of Chinese guest stars imply a maximum brightness of $<5$~mag, but more likely are brighter sightings (for details see Section \ref{sec4}). Therefore, most post-novae of a Far Eastern guest stars should still now be brighter than $18.0$~mag.

However, it is not yet known how nova remnants behave during the long time of up to $10^5$ years between two subsequent nova eruptions. \cite{duerbeck1992} determined the mean decline rate of 9 old novae to $1.0\pm0.3$ mag /century while other authors report smaller values:  for V603 Aql 1919:  $0.44\pm0.04$~mag/century \citep{johnson2014} and for  RR Pic 1925: $0.78\pm 0.20$~mag/century \citep{fuentes2018}. Most of these cases cover less than one century, they might not be valid for much larger time intervals. 

Another question arises: Is the distribution of amplitudes in Fig.~\ref{fig1} real, or biased by many unrecovered novae with amplitudes much larger than 13~mag? The total number of reported nova eruptions pre-1986 amounts to 219 events \citep{downes2005}, while the histogram in Fig.~\ref{fig1} is based on the sample of the 93 confirmed post-novae\footnote{We here exclude ancient novae like Z Cam, because they do not have an established maximum magnitude.} \citep[and references therein]{tappert2016}. Thus, about 58 per cent of the pre-1986 novae are still unidentified. If a major part of this were due to their eruption amplitude being very large, it could have significant consequences on the real amplitude distribution, i.\,e. on the shape of the histogram, especially since the latter displays a suspiciously abrupt cut-off at the 13.0 mag bin. Still, another important factor to consider is that the 126 missing post-novae will also include distant systems whose maximum brightness is already very faint, making even small-amplitude novae very difficult to detect.

In order to estimate the effect of the unidentified novae on the histogram, we analyse the part of the sample that can be considered as approximately complete for novae with amplitudes $\leq$13 mag. Medium-sized telescopes should easily detect post-novae with minimum brightness up to 20 mag photometrically, and we thus choose this value as our completeness limit. The corresponding limit for the maximum brightness of the nova eruption thus results to 7~mag. In consequence, if we extract a subsample from the pre-1986 novae defined by this limit on maximum brightness, we can assume that, in such subsample, the post-novae of all reported nova eruptions with amplitude $\leq$13 mag have been identified. The total number of objects in that subsample amounts to 60 objects, out of which 42 systems have amplitudes $\leq13.0$~mag, 10 have amplitudes $>13.0$~mag, and the remaining 8 objects are unidentified and thus have no registered minimum magnitude. To be on the safe side, we assume that in all cases the reason for the no-identification is a large amplitude, yielding an upper limit of the fraction of large-amplitude ($>13.0$~mag) post-novae of 30 per cent (18 systems). This must be compared with the observed distribution of all identified post-novae that make up the histogram in Fig.~\ref{fig1}. Here, the fraction of novae with large amplitudes amounts to only 11 per cent (10 out of 93 objects). We can thus conclude that, on the one hand, a considerable part ($\sim20$ per cent) of the yet unidentified post-novae could indeed correspond to large-amplitude objects, contributing to the $>13.0$~mag portion of the histogram in Fig.~\ref{fig1}. On the other hand, the main implication of that histogram remains valid, i.\,e. that most novae (at least 70\ per cent) have amplitudes $\leq13.0$~mag, and thus that, in general, the stellar remnant of nova eruptions observed in the pre-telescope era should be easily detectable today.

On the other hand, \cite{shara1986} suggested a ``hibernation scenario'' implying a temporary disconnection from the Roche lobe of the donor's star surface between nova eruptions. If this is valid, and if hibernation starts less than 1000 years after a nova eruption, many counterparts of older nova events could be extremely faint. The hibernation scenario would require the existence of a large number of detached red dwarf / white dwarf binaries with a configuration close to Roche lobe filling of the secondary star. Up to now, there is only one such case observed, QS Vir \citet{latkovic2019}, but, according to the \citet{gaia2018} this star is the second closest known CV-like object (distance $d = 50$~pc), apart from WZ Sge ($d = 45$~pc). \citet{henry2018} analyzed the completeness of stars or stellar systems containing white dwarfs or red dwarfs in the solar neighborhood, and concluded that a more or less complete census of such objects ($70-90$\,\%) is only achieved for distances $d = 10$~pc, while for larger distances many of those targets have escaped detection (being presently known only $\sim30$\,\% for $d = 25$~pc, $\sim15$\,\% for $d = 50$~pc, which is the distance of the nearest CVs, and only $\sim2$\,\% for $d = 100$~pc). Only very few CVs are found at these small distances, most of them are much farther away. Since detached hibernating CVs will intrinsically be much fainter than active ones, it is well possible that none of them (apart from QS Vir) has been detected until now. Therefore, any conclusion on the validity of the hibernation scenario is presently impossible.

However, the modern identifications of guest stars with ancient novae (see Table \ref{tab2}) have an average minimum magnitude of $V\approx 16.5$, not depending on the time which has passed (up to $\sim2100$ years). Therefore, we assume that the amplitude limit 13~mag, that applies to the large majority of modern classical novae (as given in Fig.~\ref{fig1}), is also valid for ancient Far Eastern nova events. Hibernation, if it exists, should become important only more than two millennia after a nova explosion.

\section{How many classical novae could have been observed by Far Eastern astronomers?}\label{sec4}
All ancient guest star records correspond to pre-telescopic surveys whose limiting magnitude for naked eye observations is of the order $V\approx 6$~mag; any classical nova detected by ancient observers should have been brighter than this (theoretical) limit. From ancient supernova records, as well as that of Mira stars, there are estimations for the real detection limit between $V\approx1.5$~mag and 3~mag in the literature.\footnote{\cite{hertzog1986} estimates 2.0~mag; \cite{clark1977} give for 3 for supernovae and for 1.5~mag for novae, both derived from Mira Ceti, \cite{strom1994} argues for 1.5~mag.} Strom (1994) mentioned the absence of early European nova detections among the guest star records between 1600 and 1900, concluding that classical novae are not present in the old guest star records. However, during the 17th century the telescope was applied to astronomical observation in Far Eastern and European astronomy and globe makers experienced much more input of newly discovered fainter stars. In 1670 a ``new star'' in Vulpecula was recognized by European astronomers, known as Nova Vulpeculae 1670 and usually identified with CK Vul. The phenomenon reveals a very peculiar reconstructed light curve with a peak magnitude 2.5 mag$ < V < 3$~mag during 10 days of the year 1671, about ten months after the original detection \citep{shara1985}. Today, CK Vul is not considered as a nova any more. Instead, it appears to have been a merger event involving a white dwarf and a brown dwarf \citep{kaminski2015}. Apart from CK Vul, there is no other confirmed case of a bright classical nova reported in Europe before 1690. It seems that the time overlap between ancient Far Eastern and modern European observations was too short to allow statistically significant results in their comparison. 

The second argument, given by \cite{strom1994}, the absence of Mira star detections among ancient records, seems not to be valid as well. In fact, Mira itself has been detected by ancient observers prior to David Fabricius' report in 1596 \citep{stephenson1976}, but on the entire sky there is no other Mira star as bright as Mira itself. There are two further Mira stars, $\chi$ Cyg and R Hya, reaching about $V\approx4$~mag at maximum light and visible from the northern hemisphere, and a few more around 5th magnitude. For $\chi$ Cyg we even found a candidate identification, the event on November 14th in 1404 CE ($\Delta\alpha = 4.8^\circ$) in the list of \cite{stephenson1976}.  According to that, any general conclusion on a limiting magnitude at 2~mag for ancient observations, based only on statistics with one (apparent) Nova case and one (or two) Mira cases, is not at all convincing. Therefore, we present our calculation in this section for three possible limiting magnitudes of ancient guest star observers, adopting the limits 2~mag, 4~mag and 5~mag (cf. Tab.~\ref{tab3} and \ref{tab4}). 

In the following paragraph, we will give a rough estimate on the total number of nova eruption events which could have been registered in pre-telescopic epochs by alert night sky observers. If all CVs are classical novae in some moment of their life we can base our calculation upon the known space density of CVs, ($\rho_{CV}$ is in the order of $10^{-5} \textrm{pc}^{-3}$, \cite{belloni2018}), their mean absolute visual magnitude at light maximum of a nova outburst (M$_{v, max} = -7.5$~mag, Warner, 1995) and the maximum distance $z_{max}$ of CVs from the galactic plane ($\pm 100$~pc, \cite{warner1995}).  Classical novae are repeating events in the same CV, with a recurrence time $T_r$ between $10^4$ and $10^5$ years \citep{yaron2005}. We adopt a guide value of $T_r = 30\,000$ years for our purpose, while the total time covered with Far Eastern guest star reports is 2222 years (from 532 BCE to 1690 CE). 

 \begin{center}
\begin{table}[t]%
\centering
\caption{The expected number of Far Eastern classical nova sightings.\label{tab3}}%
\tabcolsep=0pt%
\begin{tabular*}{20pc}{@{\extracolsep\fill}p{19ex}p{5ex}p{11ex}p{11ex}p{11ex}@{\extracolsep\fill}}
\toprule
 limiting vis. magnitude &&\textbf{2.0} & \textbf{4.0}  & \textbf{5.0}    \\
\midrule
distance $d_{max}$ of nova for absolute magnitude $-7.5 mag$ && 800 & 2000 & 3200 \\ \hline
space volume covered: $V=2\pi z_{max} d_{max}^2 / 10^9$~pc$^3$ && 0.4 &2.51 & 6.43 \\ \hline
total number $N_{CV}=V\cdot\rho_{CV}$ of cataclysmic variables in volume && 4\,020 & 25\,100 & 64\,300 \\ \hline
\multicolumn{5}{p{.97\columnwidth}}{fraction of $T_r$ covered by ancient observers between 532 BCE and 1690 CE: 2222/30000=7.4\,\%.} \\ \hline
number of nova eruptions during considered epoch $7.4\,\% \cdot N_{CV}$ && 300 & 1\,860 & 4\,760 \\
\bottomrule
\end{tabular*}
\end{table}
\end{center}

The different steps of this estimation are listed in Tab.~\ref{tab3} for three apparent magnitudes. According to this, we expect nearly 2000 sightings during the more than two millennia covered by ancient observations if a limiting magnitude of 4~mag is reached. However, the real numbers will be smaller by several reasons, not yet taken into account: It is not clear, whether really all CVs erupt as novae. The absorption by interstellar dust in the Milky Way could reduce significantly the average value of $d_{max}$ and that of volume $V$. Ancient Far Eastern observers did not have access to far southern declinations of the sky. Whenever an eruption happened near the position of the Sun (and perhaps also near the Moon), it would have escaped the attention of ancient observers. The same may have happened whenever a rapidly decaying nova (visible only for a few days) coincided with poor weather conditions. In the crowded regions of the Milky Way plane and especially near the Galactic center it could have been rather difficult to register a guest star of 4th to 5th magnitude, even for observers very familiar with the celestial constellations. 

We consider it hardly possible to estimate the influence of all these factors exactly.\footnote{Earlier scholars estimating the completeness of historical supernova observations like \cite{strom1994} chose alternative ways of figuring out the complete number which are not applicable here. \cite{duerbeck1990} estimated only the expected number from telescopic era observations and did not speculate on the completeness.}
However, even if they would reduce the above numbers by a factor 10, we still expect between $\sim30$ and $\sim480$ sightings of classical novae during the 2222 years covered by Far Eastern sources, well in accord to the existing records of ancient observations. In Table \ref{tab4} we compare these results with modern nova detection rates that coincide within the expected uncertainties, at least for the limits of 4~mag and 5~mag, while the difference at 2~mag can be explained by the very small numbers involved. It should be emphasized that most of the limitations mentioned above as possibly impeding detection are also valid for modern observers. We can conclude that a significant fraction of the ancient observations should refer to classical novae. A systematic search for their modern counterparts appears worthwhile. 

It should be emphasized that estimates given in this section suffer from many uncertainties, most of them unknown or impossible to be appraised. The only purpose of our statistics is to refer to orders of magnitude, which seem to result reasonable and compatible with the assumption that some of the oriental guest stars could refer to classical nova eruptions.

 \begin{center}
\begin{table}[t]%
\centering
\caption{Comparison to modern nova detection rates.\label{tab4}}%
\tabcolsep=0pt%
\begin{tabular*}{20pc}{@{\extracolsep\fill}p{19ex}ccc@{\extracolsep\fill}}
\toprule
 limiting vis. magnitude &\textbf{2.0} & \textbf{4.0}  & \textbf{5.0}    \\
\midrule
estimated detection rate of of novae per century from CV-density (Tab.~3) & 1.4 & 8.4 & 21.4 \\ \hline
modern detection rate according to \cite{duerbeck1990} & 6 & 13 & 18 \\\hline
modern detection rate from registered novae in the AAVSO VSX catalogue (1848--2015) & 4.2 & 10.7 & 20.8 \\  \hline
\bottomrule
\end{tabular*}
\end{table}
\end{center}

\section{On the probability of chance coincidences}\label{sec5}
Above, we state that most of the modern counterparts of Far Eastern classical nova observations should be still brighter the 18th visual magnitude. Taking into account this limit, the total numbers of CVs listed presently in the AAVSO Variable Star indeX  (VSX: AAVSO, 2005--2019; \cite{watson2006}), is 1243; in addition, we include as possible counterparts also 5 super-soft X-ray sources (type CBSS) and 179 symbiotic stars (type ZAND), revealing a total number of 1427 possible identifications.  This total number of currently known CVs with a minimum magnitude $\leq18$~mag implies that the average angular distance among them, in case of a uniform distribution on the celestial sphere, is $5.4$~degrees.  If the angular distance error is of the order of $3^\circ$, as derived in Section~2, we expect about 2 chance or random coincidences among 7 guest star identifications, based on the actual number of known cataclysmic variables and other possible identification targets. However, CVs are not distributed uniformly at the celestial sphere. Most of them are concentrated at low galactic latitudes, increasing the probability of chance coincidences there, and diminishing it for higher latitudes.

\onecolumn
\begin{figure}[t]
	\centerline{\includegraphics[width=.84\columnwidth]{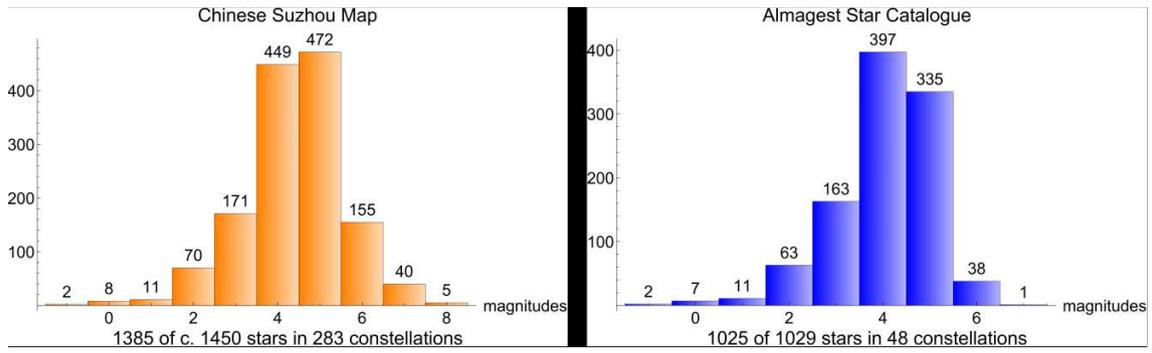}}
	\caption{Comparison of the stellar magnitude contribution in ancient western (left) and
Far Eastern (right) ancient star catalogues.\label{fig2}}
\end{figure}
\twocolumn

\section{Conclusions}\label{sec6}
Many generations of Far Eastern professional astronomers have dedicated their lives to observe the sky for more than 2000 years, recognizing and defining asterisms with many more details than the western system of constellations (cf. Fig.~\ref{fig2}). As shown in this Figure the Far Eastern catalogue has more stars and more asterisms and, thus, compiled at almost the same latitude it includes more faint stars. 

Unfortunately, the Far Eastern star catalogue is less accurate because the positions of many of the stars have not been measured. However, we hope to be able to find the modern counterparts of historical novae using historical maps and globes and using the advantage of smaller asterisms. The histogram of Chinese sources shows an impressingly large number of faint stars (magnitude 5 or fainter). Apparently, the ancient astronomers did reach these faint limits, and there is no reason to believe that they were not able to detect new guest stars down to 5th magnitude whenever it appeared in one of the well-known asterism patterns. In fact, from the reported supernova durations it was estimated that a guest star, once detected, could be discerned to $V\approx5$ or even $V\approx6$~mag (\citealt{clark1977}; \citealt{strom1994}). From these considerations, together with those given in Section~4 we can conclude that a fraction of reported guest stars should refer to classical nova eruptions. 

A successful identification of these Far Eastern classical novae with modern counterparts among cataclysmic variables should be possible under a few important conditions:

 \begin{enumerate} 
 \item Original Far Eastern sources should be re-checked, in order to assure the correct positions on the sky at which the ancient people have observed guest stars.

 \item In order to avoid a large fraction of chance coincidences, the positional uncertainties of the angular separations should be small.  

 \item The rather scarce experiences with hitherto published nova identifications suggest that (2) will be possible after application of (1): for radii of $3^\circ$ to $7^\circ$ we expect only few chance coincidences. 

 \item The typical amplitudes of classical novae imply that most of the stellar counterparts should be today at 18th magnitude or brighter. 

 \item Additional criteria not presented here (coincidences with planetary nebulae; detection of expanding nova shells etc.) will be necessary for a unique modern nova identification of an ancient event. 
 \end{enumerate}

\section*{Acknowledgments}
 N.V. and C.T acknowledge financial support from FONDECYT Regular No.~1170566, while S.H. is employed at the Friedrich-Schiller-Universität Jena. Thanks also to Monica Zorotovic and Matthias Schreiber for valuable discussions and comments, and to the Centro de Astrofísica of Universidad de Valparaíso for support. We thank the referees, Virginia Trimble and Mike Shara, for helpful comments. -- Ralph Neuh\"auser (AIU, Friedrich-Schiller-Universität Jena) had the initiative and idea to reconsider historical nova identifications including new nova candidates with terra-astronomical methods in a transdisciplinary team project. He also contributed suggestions and corrections to the preparation of the present paper, but did not agree with some approaches and conclusions presented here.

\subsection*{Author contributions}
This article summarizes mainly the preliminary study by N.V. (2014--2019) and S.H. (2017). C.T. contributed Fig.~1 and participated in all discussions. 


\nocite{*}

%

\end{document}